\documentclass[aps,prl,twocolumn,superscriptaddress,showpacs]{revtex4}
\usepackage{amsmath}
\usepackage{amssymb}
\usepackage{hyperref}
\usepackage{url}
\usepackage{graphicx}
\usepackage{dcolumn} 
\usepackage{bm}      
\usepackage{multirow}
\usepackage{epstopdf}
\usepackage{color}
\usepackage{gensymb}
\usepackage{textcomp}
\usepackage{chemformula} 

\begin{document}

\title{Charge Density Wave Order and Fluctuations above $\textit{T}_\text{CDW}$ \\ and below Superconducting $\textit{T}_\text{c}$ in the Kagome Metal \ch{CsV3Sb5}}

\author{Q.~Chen}
\affiliation{%
 Department of Physics and Astronomy, McMaster University, Hamilton, Ontario, L8S 4M1, Canada
}%
\affiliation{
 Brockhouse Institute for Materials Research, Hamilton, Ontario, L8S 4M1, Canada
}%

\author{D. Chen}
\affiliation{%
 Max Planck Institute for Chemical Physics of Solids, 01187 Dresden, Germany
}%
\affiliation{%
 College of Physics, Qingdao University, Qingdao 266071, China
}%

\author{W. Schnelle}
\affiliation{%
 Max Planck Institute for Chemical Physics of Solids, 01187 Dresden, Germany
}%

\author{C. Felser}
\affiliation{%
 Max Planck Institute for Chemical Physics of Solids, 01187 Dresden, Germany
}%
\affiliation{
 Canadian Institute for Advanced Research, Toronto, Ontario M5G 1M1, Canada
}%

\author{B. D. Gaulin}%
\affiliation{%
 Department of Physics and Astronomy, McMaster University, Hamilton, Ontario, L8S 4M1, Canada
}%
\affiliation{
 Brockhouse Institute for Materials Research, Hamilton, Ontario, L8S 4M1, Canada
}%
\affiliation{
 Canadian Institute for Advanced Research, Toronto, Ontario M5G 1M1, Canada
}%

\date{\today}

\begin{abstract}
The phase diagram of the kagome metal family \ch{AV3Sb5} (A = K, Rb, Cs) features both superconductivity and charge density wave (CDW) instabilities, which have generated tremendous recent attention.  Nonetheless, significant questions 
remain. In particular, the temperature evolution and demise of the CDW state has not been extensively studied, and little is known about the co-existence of the CDW with superconductivity at low temperatures. We report an x-ray scattering study of \ch{CsV3Sb5} over a broad range of temperatures from 300 K to $\sim$ 2 K, below the onset of its superconductivity at $\textit{T}_\text{c}\sim$ 2.9 K.  Order parameter measurements of the $2\times2\times2$ CDW structure show an unusual and extended linear temperature dependence onsetting  at $T^*$ $\sim$ 160 K,  much higher than the susceptibility anomaly associated with CDW order at $\textit{T}_\text{CDW}=94$ K.  This implies strong CDW fluctuations exist to $\sim1.7\times\textit{T}_\text{CDW}$.  The CDW order parameter is observed to be constant from $T=16$ K to 2 K, implying that the CDW and superconducting order co-exist below $\textit{T}_\text{c}$, and, at ambient pressure, any possible competition between the two order parameters is manifested at temperatures well below $\textit{T}_\text{c}$, if at all. Anomalies in the temperature dependence in the lattice parameters coincide with $\textit{T}_\text{CDW}$ for $\textit{c}(T)$ and with $T^*$ for $a(T)$.

\end{abstract}

\maketitle

\begin{figure}[tbp]
\linespread{1}
\begin{center}
\includegraphics[width=\columnwidth]{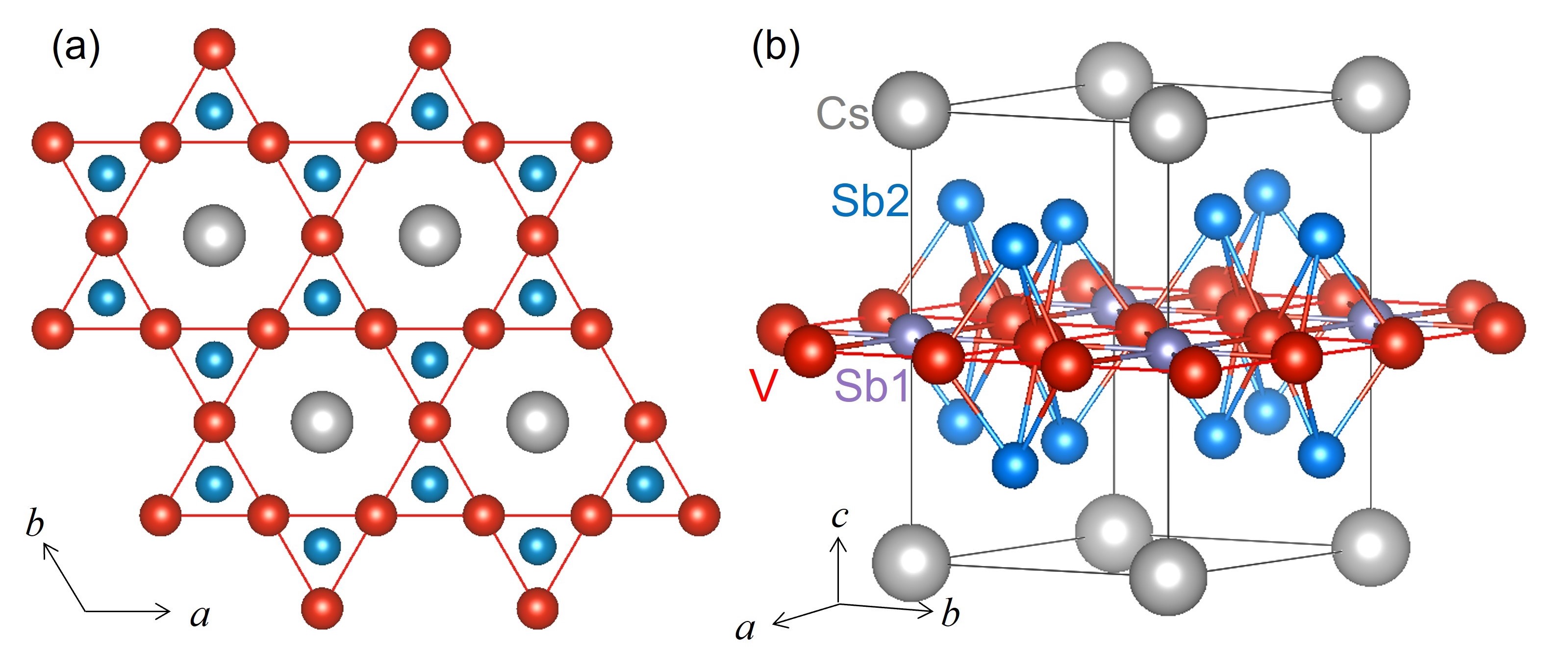}
\end{center}
\caption{The crystal structure of \ch{CsV3Sb5} is shown. a) shows a projection of the structure within the kagome plane, while b) shows a more three dimensional perspective.  Two Sb sites with different Wyckoff positions are labeled as Sb1 and Sb2.
}
\label{Fig5}
\end{figure}

The newly discovered kagome metal system \ch{AV3Sb5} (A = K, Rb, Cs) \cite{19ortiz} has been extensively studied in the last two years due to the rich cooperative physics that they display. They represent a new class of topological kagome metal with both superconducting and charge density wave (CDW) instabilities \cite{20ortiz,21hchen,21denner,21ortiz,21yin,21yjiang,21liang}.  These compounds appear to be nonmagnetic and undergo CDW transitions at relatively high temperatures, $\textit{T}_\text{CDW}=78$ K, 104 K and 94 K respectively, before  displaying superconductivity at low temperatures, with $\textit{T}_\text{c}=0.93$ K, 0.92 K and 2.5 K respectively \cite{19ortiz,20ortiz,21ortiz}. Their layered crystal structure is based on a stacking of A-Sb2-VSb1-Sb2-A layers (see Fig.~\ref{Fig5}(b)) with hexagonal symmetry in the space group $P6/mmm$, and a kagome network of vanadium atoms coordinated by antimony in the basal plane (Fig.~\ref{Fig5}(a)). The vanadium and antimony layers are separated by alkali layers and form the quasi-2D structure \cite{21wilson}. This system is known to provide a platform for a variety of collective quantum phenomena, including a large anomalous Hall effect \cite{21yu,21dchen}, a nontrivial $\mathbb{Z}_{2}$ topology \cite{21ortiz}, superconductivity and chiral charge ordering \cite{21shumiya,21lin}, and all of this is motivating their great current interest. 

Focusing on the case of \ch{CsV3Sb5}, several recent studies have investigated its electronic ordering instability, a CDW that is associated with a magnetic susceptibility anomaly at $\textit{T}_\text{CDW}=94$ K \cite{21ratcliff,22wu,21stahl,22liu}. Somewhat surprisingly the CDW  order in \ch{CsV3Sb5} is three-dimensional \cite{21liang,21ortiz,21hchen,21hzhao,21zwang,21wang,21lih}, which means that there are electron correlations not only in the kagome plane (\textit{ab}-plane), but also between the kagome layers, along the \textit{c}-axis. However, a consensus on the details of the structure of the CDW ground state and the mechanism for its formation is currently lacking. While the in-plane superstructure is known to be $2\times2$, the out-of-plane superstructure component remains controversial and not fully characterized. For example, Liang \textit{et al} demonstrated a $2\times2\times2$ CDW state using scanning tunneling microscopy techniques \cite{21liang} while the Ortiz and Wilson group results have indicated a $2\times2\times4$ state at 15 K by surveying the reciprocal space with high resolution synchrotron x-ray diffraction \cite{21ortiz}. Moreover, a recent x-ray study by Stahl and collaborators has reported a measurement of the order parameter of several CDW Bragg intensities, suggesting a $2\times2\times2$ superstructure below 60 K, with a transition to $2\times2\times4$ above 60 K and below 94 K \cite{21stahl}. A successive high-resolution x-ray work found different results below 60 K and claimed the coexistence of $2\times2\times2$ and $2\times2\times4$ CDW stacking phases which compete each other \cite{22xiaoq}. Another recent work combined high-resolution x-ray diffraction, scanning tunneling microscopy and scanning transmission electron microscopy and suggested that the distinct $1\times4$ phase emerges uniquely on the surface \cite{21lihx}. The detailed CDW structure thus remains an open question.

In this letter, we report a temperature dependent x-ray diffraction study of the CDW instabilities in high-quality single crystal \ch{CsV3Sb5}, over a broad range of temperatures from 300 K to $\sim$ 2 K, below the onset of superconductivity in our single crystal of \ch{CsV3Sb5} at $T_c\sim$ 2.9 K. Our measurements show CDW superlattice peaks at [1.5, 1.5, $\pm0.5$] with a relatively strong temperature dependence, as well as broad CDW peaks at [1.5, 1.5, -0.25] and [1.5, 1.5, 0] without.  These correspond to  $2\times2\times2$, $2\times2\times4$, and $2\times2\times large$ CDW states in the bulk of \ch{CsV3Sb5}, respectively.

Our focus is on the $2\times2\times2$ CDW instability which shows strong temperature dependence.  Surprisingly, its order parameter extends beyond $\textit{T}_\text{CDW}=94$ K in \ch{CsV3Sb5}, to $T^*$ $\sim$ 160 K, a temperature scale that has not been previously discussed for this material.  We also measured the temperature dependence of the \ch{CsV3Sb5} crystal lattice parameters from 2 K to 300 K which will be discussed below.

High-quality single crystals of \ch{CsV3Sb5} were grown using the self-flux method that has been reported elsewhere \cite{20ortiz}. Magnetization measurements were performed using a Quantum Design superconducting quantum interference device magnetometer (SQUID) in reciprocating sample option (RSO) mode. Heat capacity measurements were performed with the relaxation method using the Quantum Design physical properties measurement system (PPMS) HC option.

Our x-ray scattering studies of these single crystals were performed using a Huber four circle diffractometer with a closed cycle refrigerator equipped with a Joule-Thompson stage. X-ray scattering measurements were performed using Cu $K\alpha$ radiation ($\lambda=1.54~\text{\AA}$) produced by an 18 kW rotating anode source with a vertically focused pyrolytic graphite monochromator. These x-ray measurements are quite distinct from high resolution synchotron x-ray measurements, which typically employ a small spot size and experience significant beam heating of the samples.  The present measurements were performed in transmission geometry and employed a beam covering most of the 3 mm $\times$ 3 mm $\times$ 20 $\mu$m single crystal sample.  The large beam size allows relaxed resolution appropriate to better capture diffuse scattering.  The relatively low power and distributed beam results in minimal beam heating and equilibrated sample temperatures to as low as 2 K. This experimental configuration is ideal for x-ray measurements at very low temperatures. Measurements of fluctuations down to a minimum temperature of 300 mK have been successfully performed in other systems in this manner \cite{07ruff}.

\begin{figure}[tbp]
\linespread{1}
\begin{center}
\includegraphics[width=\columnwidth]{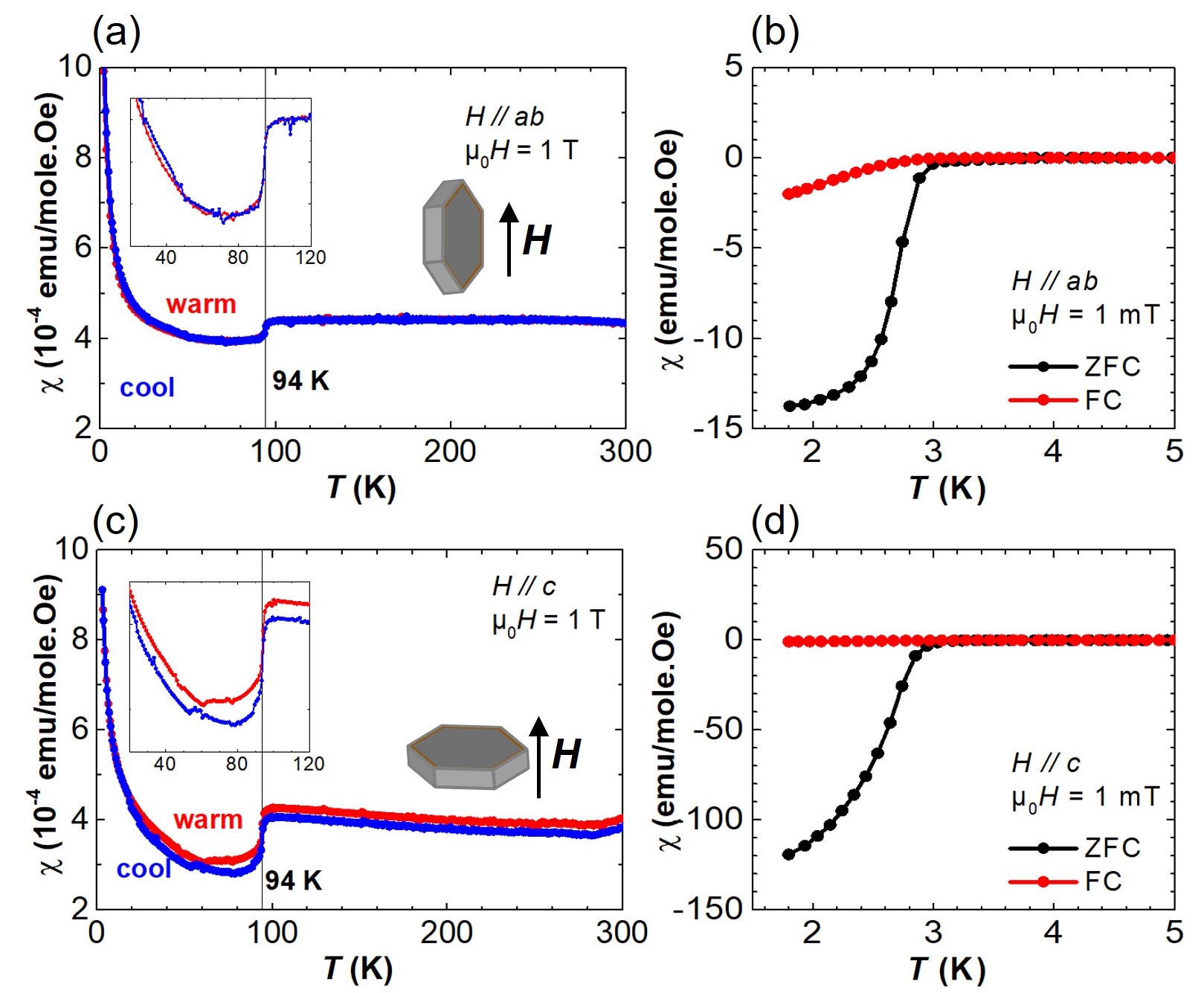}
\end{center}
\caption{Magnetic susceptibility measurements for single crystals of \ch{CsV3Sb5} with an external magnetic field applied within the kagome plane (a, b) and perpendicular to it (c, d) are shown. (a, c) Magnetic susceptibility measurements with $\mu_0H=1$ T, show an anomaly at $\textit{T}_\text{CDW}=94$ K. The inset shows another anomaly at $\sim60$ K in (c) only, field out of plane. (b, d) Susceptibility measurements at low temperatures using a small dc field $\mu_0H=1$ mT, show the onset of superconductivity below $\textit{T}_\text{c}=2.9$ K.
}
\label{Fig1}
\end{figure}


Magnetization measurements were performed both with the external magnetic field applied within the kagome plane ($H \parallel ab$, IP), Fig.~\ref{Fig1}(a)(b), and out of the plane ($H \parallel c$, OP), Fig.~\ref{Fig1}(c)(d). Measurements were performed at both high and low temperature in order to characterize the CDW and the superconducting transitions. The measurements to high temperature in Fig.~\ref{Fig1} (a) and (c) were performed  with a 1 T external magnetic field from 2 K to 300 K while the low-temperature measurements employed a field of 1 mT from 1.8 K to 5 K using the zero-field-cooled (ZFC) warm-up, followed by field-cooled (FC) cool-down protocols. The sharp transition observed in the susceptibility data at $\textit{T}_\text{CDW}=94$ K (Fig.~\ref{Fig1}(a)(c)) agrees well with previous reports \cite{20ortiz,21wang,21zwang} and corresponds to the CDW transition.

Low temperature susceptibility measurements are shown in Fig.~\ref{Fig1}(b)(d). The sudden drop of the ZFC signal corresponds to the superconducting transition. We note that our superconducting transition temperature so-identified is $\textit{T}_\text{c}$ $\sim$2.9 K, slightly higher than those previously reported for \ch{CsV3Sb5}, which varied between 2.5 K and 2.8 K \cite{20ortiz,21hchen,21fyu}. We attribute the difference to the high quality of our as-grown \ch{CsV3Sb5} sample \cite{21hchen}.
Although the $\textit{T}_\text{CDW}$ and $\textit{T}_\text{c}$ identified with field IP  and OP  measurements are the same, the magnitudes of the susceptibilities are different, showing significant anisotropy, as expected given the quasi-two-dimensional nature of this kagome metal \cite{19ortiz,20ortiz,21jiangk,21wilson}. A further anisotropic feature observed in the susceptibility is at $\sim$ 60 K, where there is a weaker but sharp anomaly observed in the OP measurement, shown in panel (c), and this feature seems to be absent in the IP measurement shown in panel (a). It also shows a small thermal hysteresis between the warming and cooling measurements which indicates a coupling to the lattice. This feature at $\sim$ 60 K has not been observed in previously reported susceptibility data \cite{20ortiz,21wang}. However, previous x-ray \cite{21stahl}, magnetotransport \cite{21hchen,21xiangy}, STM \cite{21hzhao,21liang}, coherent phonon spectroscopy \cite{21ratcliff, 21wangzx,21wuq}, and muon spin relaxation \cite{21yul} measurements have all found anomalies at this temperature, and it has been argued to be due to the 1$\mathbf{Q}$ unidirectional nature of the CDW.  The observation of this feature in the OP ($H \parallel c$) susceptibility measurement alone would seem to be consistent with this argument. Nonetheless, a recent polarization-resolved electronic Raman spectroscopy study as well as x-ray results combined with the group theoretical analysis suggest that the uniaxial charge modulation below 60 K may not be a bulk effect \cite{22wu}. 

\begin{figure}[tbp]
\includegraphics[width=\columnwidth]{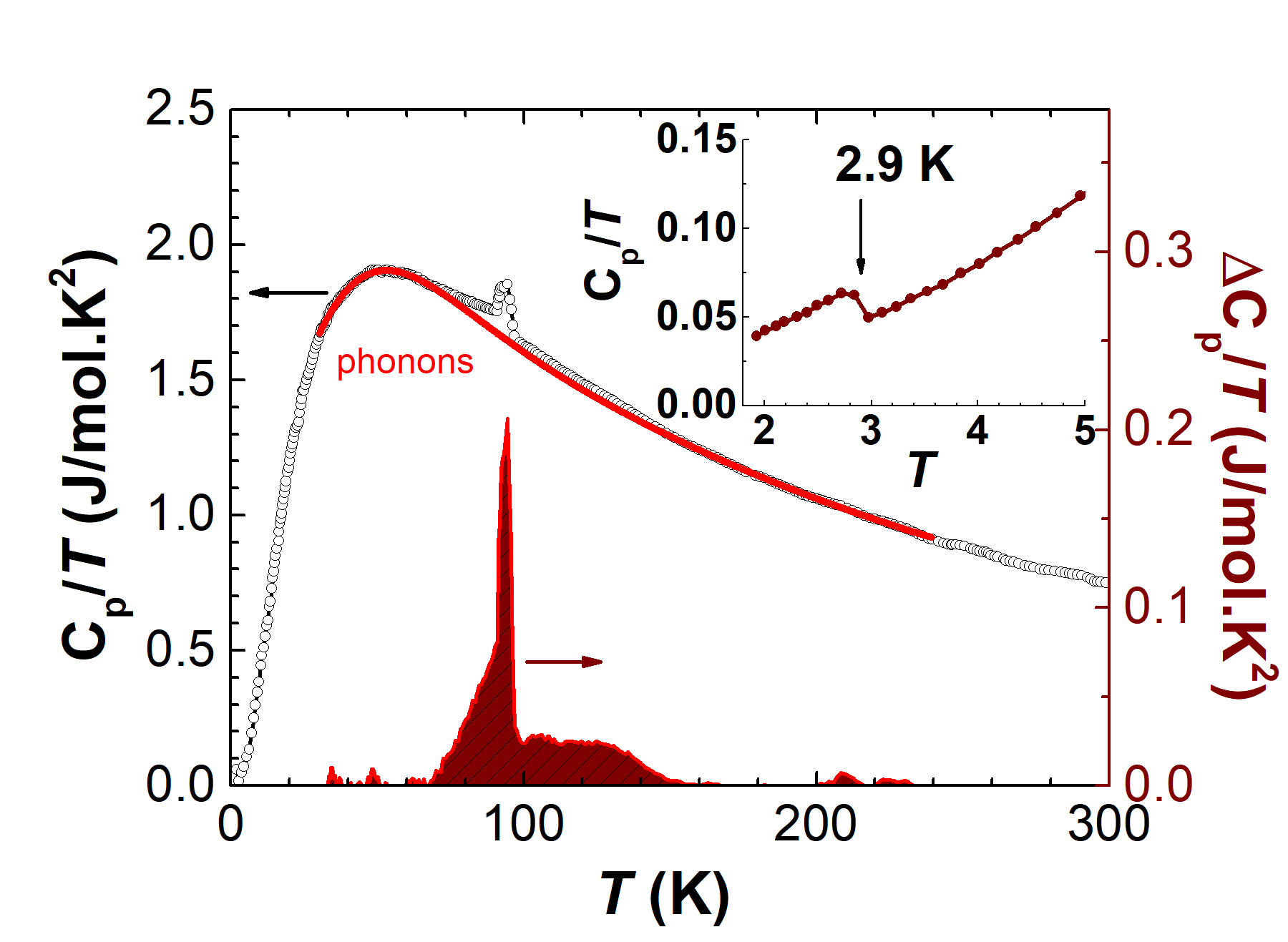}
\caption{
The temperature dependence of the specific heat $C_p$ measured at zero field shows a clear transition at $\textit{T}_\text{CDW}=94$ K. The red curve shows a fit to the lattice contribution of $C_p$ using a Thirring model. The subtracted data $\Delta C_p$ is shown at the bottom with its scale on the right y-axis. The inset depicts the low-temperature regime to $C_p$, highlighting the onset of superconductivity below $\textit{T}_\text{c}=2.9$ K.
}
\label{Fig6}
\end{figure}

Specific heat measurements at zero field also show a sharp transition at $\textit{T}_\text{CDW}=94$ K, as Fig.~\ref{Fig6} demonstrates for our single crystal. The inset depicts the low-temperature regime data to highlight the onset of superconductivity below $\textit{T}_\text{c}=2.9$ K, which agrees well with our magnetic susceptibility measurements above. The red curve (labeled phonons) is our best fit to the high temperature lattice contribution of \ch{CsV3Sb5}, approximated by the Thirring model \cite{13_thirring,89_thirring, 05_thirring},
which allows us to preferentially assess the anomalous contribution above $\textit{T}_\text{CDW}$.
The subtracted data $\Delta C_p/T$, 
is shown at the bottom of the figure, with its scale on the right. We note that, it is difficult to separate the CDW contribution over this whole temperature range, it is clear that the CDW entropy release is strongly peaked at $\textit{T}_\text{CDW}=94$ K, 
perhaps due to a weak first-order character arising from weak coupling between electronic and lattice degrees of freedom.  It is also clearly
extended across a broad range of temperatures up to $\sim$150 K.  As will be discussed, this extended temperature dependence of the CDW $\Delta C_p/T$ above $\textit{T}_\text{CDW}=94$ K is consistent with our x-ray scattering results. 


\begin{figure*}[tbp]
\linespread{1}
\begin{center}
\includegraphics[width=6in]{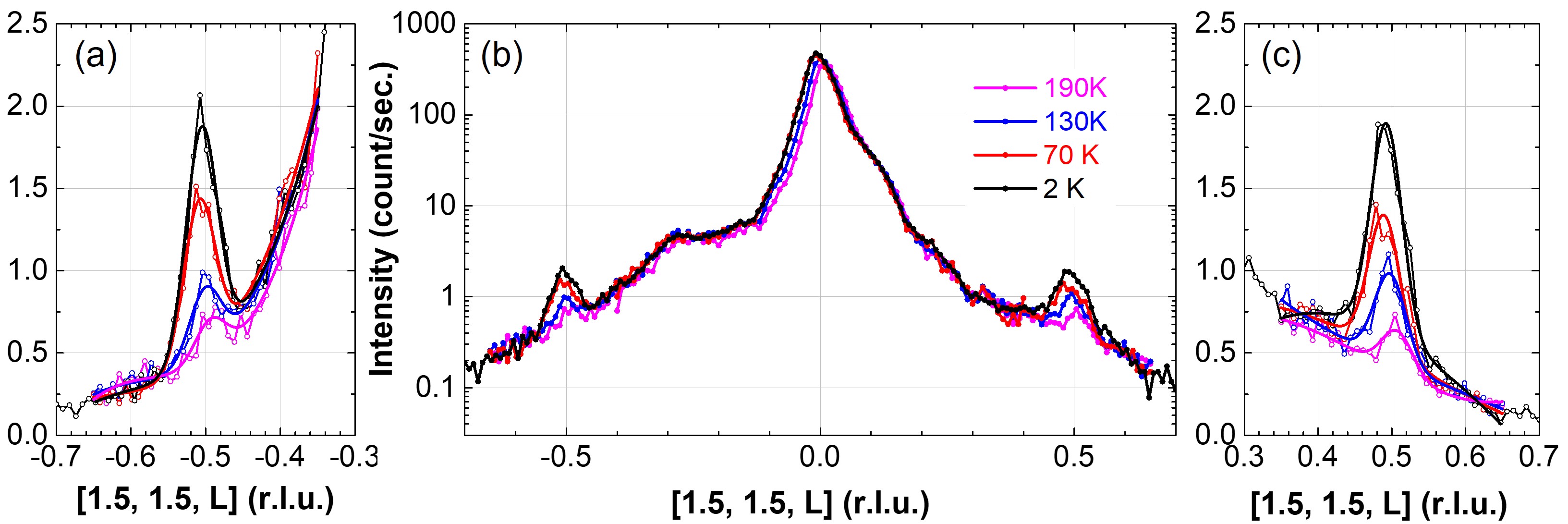}
\end{center}
\caption{(color online) The temperature dependence of the x-ray diffraction intensity (normalized to counts per second) at [1.5, 1.5, \textit{L}] is shown.  Panel (a) and (c) show the CDW peaks at $L=-0.5$ and $L=0.5$, respectively, and intensity is plotted on a linear scale. Panel (b) shows the the full [1.5, 1.5, \textit{L}] scans, with intensity plotted on a log-scale. The temperature-dependent peaks at half-integer reflections [1.5, 1.5, $\pm$0.5] are due to the $2\times2\times2$ CDW structure.
}
\label{Fig2}
\end{figure*}

We next investigate the CDW state by x-ray diffraction over a broad temperature range from 300 K down to 2 K.  In Fig.~\ref{Fig2} we show the main results of this x-ray scattering study, which are line scans in reciprocal space carried out along [1.5, 1.5, $L$] at different temperatures. Data at four representative temperatures are shown in this figure. We observe significant temperature dependence in this scan only at $L = \pm0.5$, consistent with a $2\times2\times2$ CDW state. A strong temperature-independent peak at $L = 0$, is also evident in Fig.~\ref{Fig2}(b). In Fig.~\ref{Fig2}(a) and (c), the [1.5, 1.5, $\pm0.5$] superlattice reflections are fit to a Gaussian function with a fixed full-width-at-half-maximum (FWHM) of 0.05 $r.l.u.$ (reciprocal lattice unit), displayed with like-color curves on top of the data.

\begin{figure}[tbp]
\linespread{1}
\begin{center}
\includegraphics[width=\columnwidth]{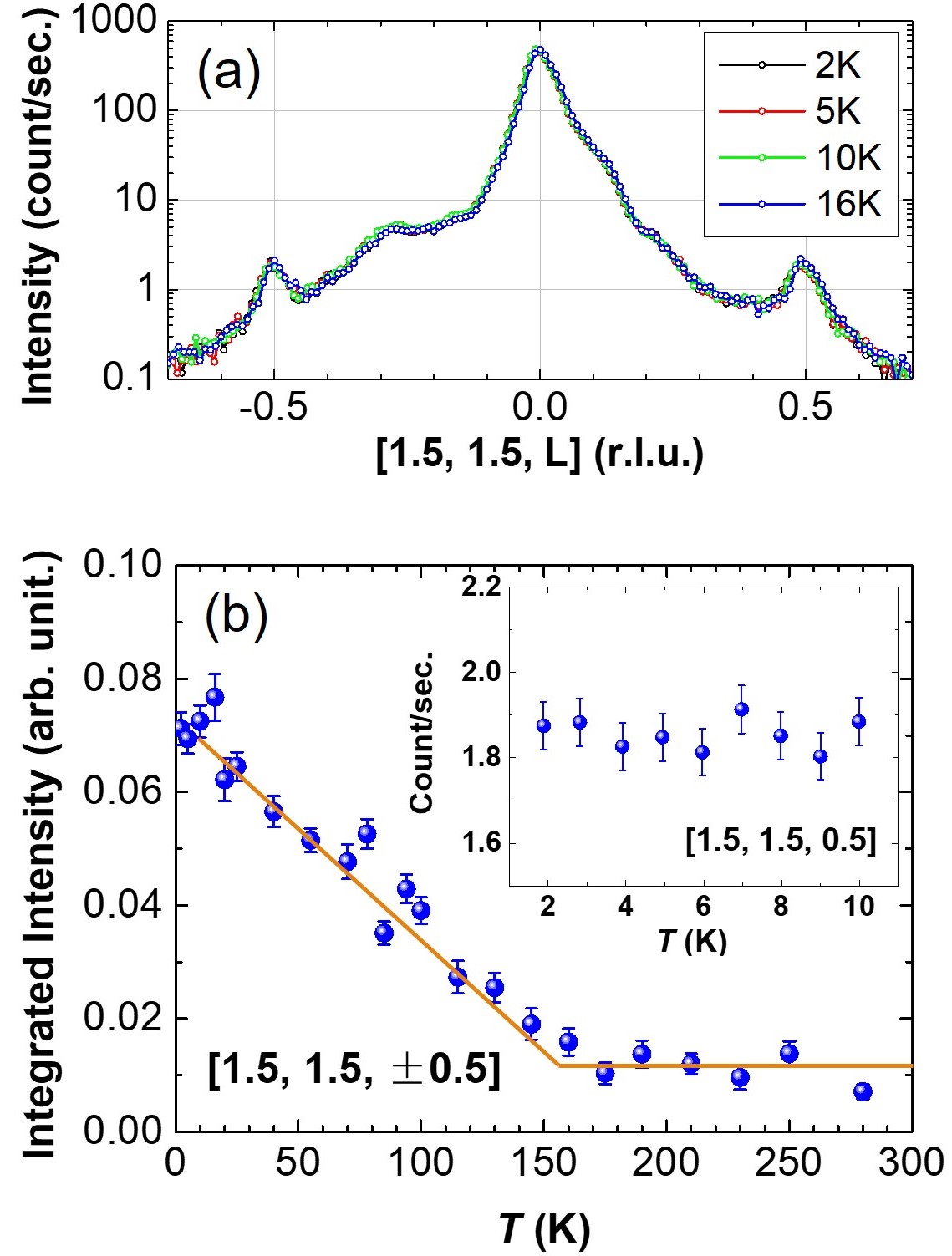}
\end{center}
\caption{(color online) (a) The temperature dependence of the x-ray diffraction scans across [1.5, 1.5, \textit{L}] is shown for temperatures below 20\,K, and with intensity on a logarithmic scale. (b) The temperature dependence of the integrated peak intensity at the [1.5, 1.5, $\pm$0.5] CDW reflections, the order parameter, is shown. The inset shows the peak intensity at [1.5, 1.5, 0.5] at low temperatures, from 2 K to 10 K.}
\label{Fig3}
\end{figure}

We attribute the temperature dependence at [1.5, 1.5, $\pm0.5$] superstructure reflections to the $2\times2\times2$ CDW order and its fluctuations.  The extracted integrated intensities arising from the Gaussian fittings are shown as an order parameter in Fig.~\ref{Fig3}(b). It's inset shows the peak intensity of the [1.5, 1.5, 0.5] superstructure reflection at the lowest temperatures, ranging from 2 K to 10 K, where it is clear that it is almost temperature-independent. To confirm this low temperature behaviour, we performed careful line scans along [1.5, 1.5, $L$] at 2 K, 5 K, 10 K and 16 K as shown in Fig.~\ref{Fig3}(a), which show no appreciable temperature dependence at these low temperatures.

Previous high-pressure measurements on this kagome metal system \ch{AV3Sb5} show that superconductivity co-exists with CDW order under ambient pressure, however superconducting $\textit{T}_\text{c}$ increases while $\textit{T}_\text{CDW}$ decreases with increasing pressure, suggesting a competition between CDW order and superconductivity \cite{21fyu,21wilson,21songy,21wangn}.
Evidence for such competition is also provided by hole-doping studies\cite{21songy,21liuy,22oey}, systematic studies of finite crystal thickness \cite{21songy,21songb}, and measurements under uniaxial strain \cite{21qiant}.  In contrast, our x-ray results on the ambient pressure CDW state in \ch{CsV3Sb5} show it to be robust and to co-exist well with its superconducting instability below $\textit{T}_\text{c}=2.9$ K, at least as low as $T=2$ K.

Our x-ray diffraction order parameter for the CDW peak at [1.5, 1.5, $\pm0.5$] shows several intriguing features, in addition to its co-existence with superconductivity at low temperatures.  First, Fig.~\ref{Fig3}(b) shows the CDW order parameter to extend in temperature to $\sim160$ K, more than 1.5 times of the $\textit{T}_\text{CDW}=94$ K determined from the magnetic susceptibility and specific heat measurements.  Secondly, it possesses a remarkably linear temperature dependence over an extended temperature regime.  Earlier high-resolution x-ray measurements have reported related CDW Bragg scattering to turn on below $\textit{T}_\text{CDW}=94$ K  \cite{20ortiz,21ortiz,21stahl,21lihx}.  Our new results imply that the order parameter above $\textit{T}_\text{CDW}=94$ K originates from fluctuations, rather than true long range order.  However it also implies that these fluctuations are strong.  A similar conclusion is reached for the strong, temperature-independent peak at [1.5, 1.5, 0]. This has not been reported previously, and its presence would again argue for strong fluctuations related to the CDW state at temperatures high compared to $\textit{T}_\text{CDW}=94$ K.

Of course, critical fluctuations are routinely measured in order parameter studies of continuous phase transitions.  However, for three dimensional (3D) materials, strong critical fluctuations are typically limited to temperatures within $\sim$ 0.2 in $(T-T_\text{c})/T_\text{c}$.  Low dimensional systems, in contrast, are known to exhibit extended critical scattering, such that fluctuations developed well above the critical temperature, $T_\text{c}$, but only lock into a 3D structure below $T_\text{c}$.  

Strong fluctuation effects associated with the order parameter have been reported in another \textit{d}1 electronic system, \ch{TiOCl}, under similar low resolution diffraction conditions as those which are relevant here \cite{08clancy,05hemberger}.  In \ch{TiOCl}, the structural distortion arises from spin-Peierls phenomena, and it displays both a commensurate and incommensurate dimerized structure.  However, the measured commensurate order parameter extends well above the transition temperature and displays a growth which is linear in decreasing temperature, very reminiscent of that observed here for \ch{CsV3Sb5}. The entropy release of \ch{TiOCl} to $\sim150$ K, that is $\sim1.6\times$ its highest temperature $T_\text{c2}$ \cite{05hemberger}, is also very similar to what we observed here in Fig.~\ref{Fig6}.

\begin{figure}[tbp]
\linespread{1}
\begin{center}
\includegraphics[width=\columnwidth]{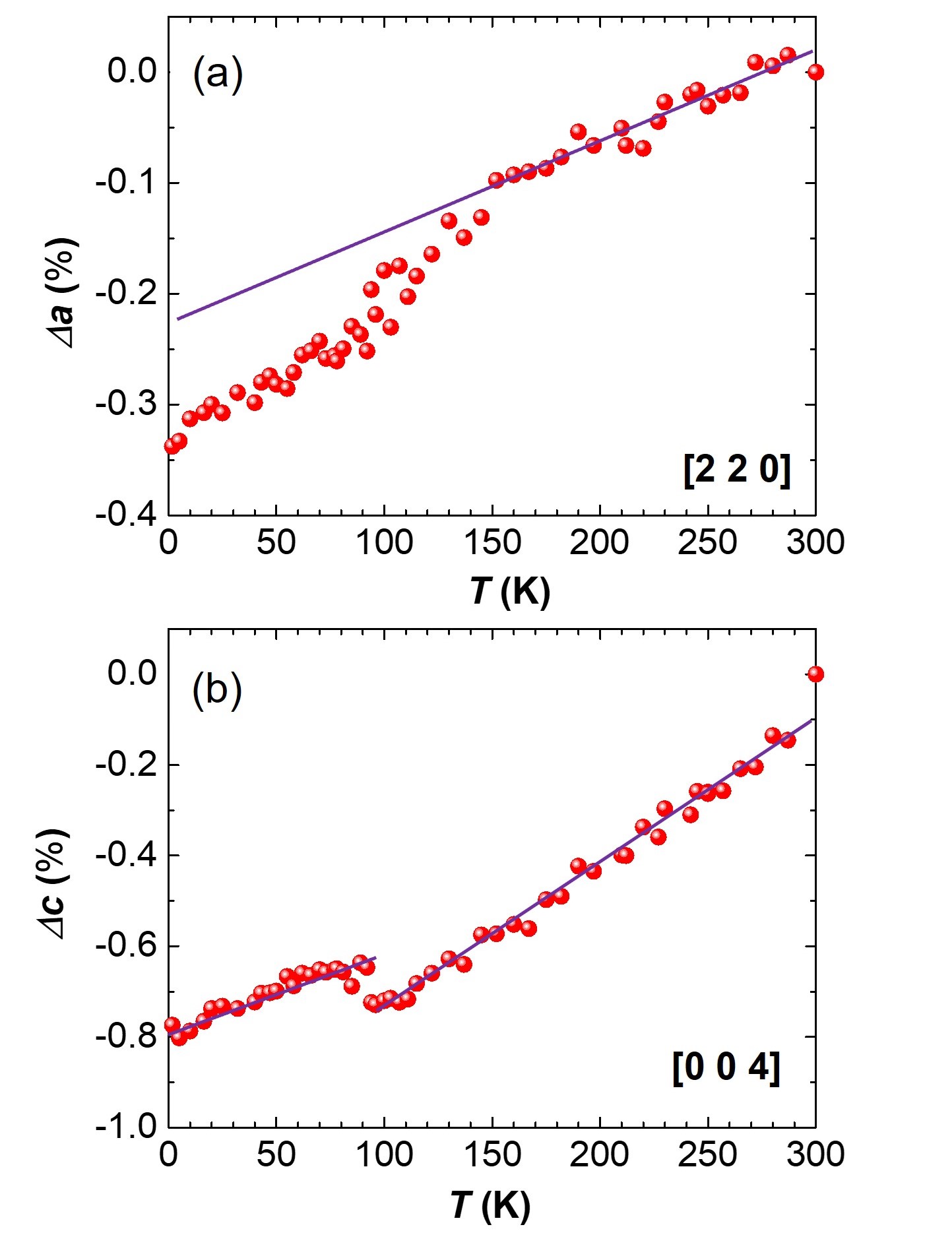}
\end{center}
\caption{The relative change in the lattice constants as a function of temperature in \ch{CsV3Sb5} are shown. (a) The \textit{a} lattice constant shows a slope change at temperatures below $T^*$ $\sim$160 K, and perhaps a second one below $\textit{T}_\text{CDW}=94$ K. (b) The \textit{c} lattice constant shows an abrupt change roughly coincident with $\textit{T}_\text{CDW}=94$ K. All straight lines are intended as guides-to-the-eye.
}
\label{Fig4}
\end{figure}

We have also measured the temperature dependence of the lattice constants for single crystal \ch{CsV3Sb5} by tracking the peak positions of the [2, 2, 0] and [0, 0, 4] reflections as a function of temperature from 2 K to 300 K. The temperature dependence of the lattice constants \textit{a} and \textit{c} are plotted in Fig.~\ref{Fig4}, where the relative change is normalized to the room temperature data (300 K).

The \textit{c} lattice constant, along the stacking direction, shows an abrupt change near $\sim90$ K, which is consistent with the first-order-like anomaly in the susceptibility associated with the CDW transition temperature $\textit{T}_\text{CDW}=94$ K, shown in Fig.~\ref{Fig1}(a) and (c).  In contrast, the \textit{a} lattice constant shows a change of slope near $T^*$ $\sim160$ K, and perhaps a second one near $\textit{T}_\text{CDW}=94$ K.  These sets of lattice parameter measurements were performed under both warming and cooling protocols, and no obvious hysteretic behavior was observed.


To conclude, we present detailed x-ray diffraction studies of the CDW state in the bulk of \ch{CsV3Sb5} across a very broad range of temperatures.  It reveals a remarkable linear order parameter associated with its $2\times2\times2$ CDW superlattice structure, which extends well above $\textit{T}_\text{CDW}=94$ K.  Above $\textit{T}_\text{CDW}$, we associate it with strong CDW fluctuations which eventually merge into background at $T^* \sim 160$ K, close to 1.7 $\times$ $\textit{T}_\text{CDW}$.  Our low temperature x-ray diffraction measurements down to 2 K, suggest that the CDW state in \ch{CsV3Sb5} remains robust on entering the superconducting state below $\textit{T}_\text{c}=2.9$ K. Finally, we report the relative change in lattice constants as a function of temperature in \ch{CsV3Sb5}, where a slope change of the in-plane lattice parameter \textit{a} is observed at $\sim160$ K, and a discontinuity in the \textit{c} lattice parameter is observed near $\textit{T}_\text{CDW}=94$ K.

We acknowledge useful conversations with J. P. Clancy. This work was supported by the Natural Sciences and Engineering Research Council of Canada.

\bibliography{CsV3Sb5}

\end{document}